# Tunable stacking fault energies by tailoring local chemical order in CrCoNi medium-entropy alloys

**Jun Ding[1], Qin Yu[1], Mark Asta[1,2*], Robert O. Ritchie[1,2†]**

[1] Materials Sciences Division, Lawrence Berkeley National Laboratory, Berkeley, CA94720, USA
[2] Department of Materials Science & Engineering, University of California, Berkeley, CA 94720, USA

To whom correspondence may be addressed:

Mark Asta, Department of Materials Science & Engineering, University of California, Berkeley, CA 94720
Tel: (530) 848-9881; email: mdasta@berkeley.edu

Robert O. Ritchie, Department of Materials Science & Engineering, University of California, Berkeley, CA 94720
Tel: (510) 409-1779; email: roritchie@lbl.gov

**Keywords:** Medium-entropy alloys; local chemical order; stacking fault energy; transformation-induced plasticity

**High-entropy alloys (HEAs) are an intriguing new class of metallic materials due to their unique mechanical behavior. Achieving a detailed understanding of structure-property relationships in these materials has been challenged by the compositional disorder that underlies their unique mechanical behavior. Accordingly, in this work, we employ first-principles calculations to investigate the nature of local chemical order and establish its relationship to the intrinsic and extrinsic stacking fault energy (SFE) in CrCoNi medium-entropy solid-solution alloys, whose combination of strength, ductility and toughness properties approach the best on record. We find that the average intrinsic and extrinsic SFE are both highly tunable, with values ranging from -43 mJ.m$^{-2}$ to 30 mJ.m$^{-2}$ and from -28 mJ.m$^{-2}$ to 66 mJ.m$^{-2}$, respectively, as the degree of local chemical order increases. The state of local ordering also strongly correlates with the energy difference between the face-centered cubic (*fcc*) and hexagonal-close packed (*hcp*) phases, which affects the occurrence**

**of transformation-induced plasticity. This theoretical study demonstrates that chemical short-range order is thermodynamically favored in HEAs and can be tuned to affect the mechanical behavior of these alloys. It thus addresses the pressing need to establish robust processing-structure-property relationships to guide the science-based design of new HEAs with targeted mechanical behavior.**

High-entropy alloys (HEAs), also referred to as multi-principal element alloys, have emerged as an exciting new class of metallic structural materials. These alloys are, in principle, single-phase crystalline solid solutions comprising multiple elements, typically in equal molar ratios (1-5). The original rationale guiding the discovery of HEAs was that the configurational entropy contribution to the total free energy in multi-component concentrated alloys can stabilize the solid-solution state relative to a multi-phase microstructure composed of chemically-ordered intermetallic phases (1,2). Currently, HEAs are attracting extensive research interest because some of these systems, in particular those based on the *fcc* CrCoNi system, have been found to display excellent mechanical properties, including very high fracture toughness and high strength (6,7). The five-element (so-called Cantor) HEA, CrMnFeCoNi, was the first to be shown to exhibit exceptional damage tolerance, which can be further enhanced at cryogenic temperatures (6). Additionally, the equiatomic, three-element (medium-entropy) CrCoNi alloy has been observed to display even better properties; at 77 K, this alloy displays tensile strengths of 1.4 GPa, tensile ductilities of ~90%, and fracture toughness values of 275 MPa√m, approaching the best damage-tolerance on record (7). A synergy of deformation mechanisms appears to be the basis of the exceptional mechanical behavior of these alloys (8-13). For instance, in the stronger medium-entropy CrCoNi alloy, nanotwinning occurs at room temperature and leads to the formation of a hierarchical twin network, where the twin boundaries act as barriers to dislocation motion, providing for strength, whereas both full and partial dislocations can move rapidly along the boundaries themselves for ductility (11). Additionally, a lamellar *hcp* phase has been reported to form in this alloy at higher strain levels (13).

Such unique mechanical behavior for the CrCoNi and CrMnFeCoNi alloys appears to be associated with their low stacking fault energies (SFE), as the SFE is one of the most significant parameters influencing plastic deformation, dislocation mobility, deformation twinning as well as occurrence of phase transformations in conventional crystalline alloys (14-16) (see the SI Text for further discussions). For the CrCoNi alloy, experimental measurements of the separation of partial dislocations suggested values of the SFE of 22±4 mJ.m$^{-2}$ (9); conversely, first-principles density-functional-theory (DFT) calculations have yielded *negative* values of the SFE for this alloy, *e.g.*, -24 mJ.m$^{-2}$ (11), -62 mJ.m$^{-2}$ (17) or -40 mJ.m$^{-2}$ (18), which are apparently at odds with the measured finite dislocation dissociation widths. Given the importance of the SFE in governing the nanoscale mechanisms active during plastic deformation, this discrepancy between measurement and computation clearly warrants further investigation.

The unique properties of HEAs mainly originate from their multi-components and special chemical structure, compared to conventional crystalline alloys. In HEAs, the atomic structure is characterized by an underlying crystal lattice (*e.g.*, *fcc*, *bcc* or *hcp*), but with atoms of different elements occupying the sites of the lattice in a non-periodic manner. As illustrated schematically in Fig. 1, on one hand, the distribution of heterogeneous local environments in HEAs should be broader than that of conventional crystalline alloys that display a well-defined "host" lattice set by the majority chemical species (solvent) with discrete defects associated with the (solute) alloying additions; on the other hand, the local environment of HEAs also differ from that in metallic glasses, which display amorphous structures with varying degrees of short-to-medium-range structural and chemical order, as illustrated by the even wider and continuous distribution (19).

In the simplest description, the local chemical environments of HEAs can be considered as representing a random distribution of different atomic species over crystal lattice sites, *i.e.*, the maximum configurational entropy state (1). However, as has been discussed in the literature (20-23), this description is likely to be oversimplified, as HEAs may display significant local chemical order, *i.e.*, ordering in the first few neighbor atomic shells to enhance the number of energetically preferred bond types. This raises a salient question: *To what degree does local chemical order exist in HEAs and how could it affect their deformation mechanisms and mechanical properties?* The question is not only important for a more complete fundamental understanding of mechanical behavior, but also for practical applications, since the state of local chemical order can be tuned by varying materials processing and alloy composition. Accordingly, in this work we perform detailed first-principles calculations to elucidate the relative energetics of planar faults governing slip in ternary solid-solution CrCoNi medium-entropy alloys, with a focus on understanding how chemical short-range order (SRO) in these multicomponent solid solutions can affect SFEs and their role in governing deformation mechanisms.

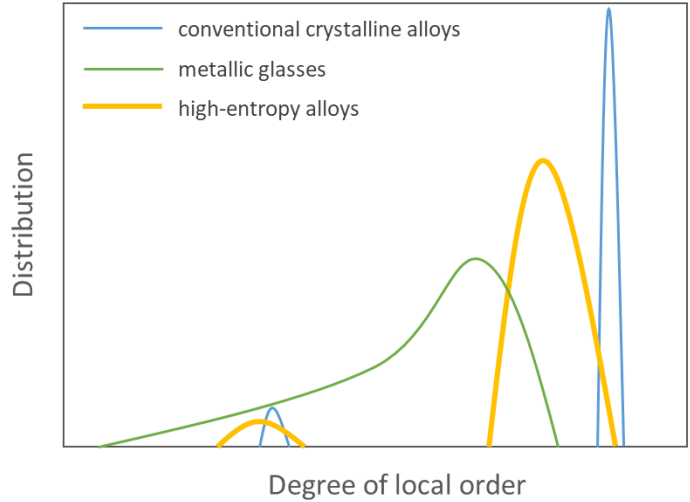

**Figure 1. Description of local environment in high-entropy alloys.** Schematic illustration contrasting the distribution of local order (both structural and chemical) in HEAs, as compared with two other alloy systems, specifically metallic glasses and conventional polycrystalline alloys. The two crystal systems both have bifurcated local environments composed of a near-perfect lattice (well-defined spikes), plus discrete defects such as dislocations and vacancies (small peaks), but the distribution for HEAs is much boarder due to their disordered chemical structure. In contrast, the structure of metallic glasses features varying degrees of short-to-medium-range icosahedral order that is reflected in their wide and continuous distribution (19).

## Results

**Local chemical order in CrCoNi solid-solution alloys**

For the purpose of examining computationally the effect of local chemical order on the stacking fault energies of CoCrNi alloys, it is necessary to develop realistic models of chemical SRO in this system. Previous experimental studies employing Extended X-Ray Adsorption Fine Structure (EXAFS) measurements (21) have shown that this SRO is characterized by a preference for Cr atoms to bond with unlike atoms (Co and Ni). Although these experiments clearly establish the presence of measurable chemical SRO in the alloys, they do not provide sufficient information to unambiguously develop atomistic models. In earlier computational work (20), simulations employing a DFT-based lattice Monte-Carlo (MC) approach established chemical SRO in qualitative agreement with experimental measurements. We thus employ a similar approach using larger systems with a larger number of independent samples to develop models for CoCrNi solid solutions with varying degrees of chemical SRO. Specifically, we have investigated local chemical order in *fcc* CrCoNi solid-solution alloys using DFT-based MC simulations as described in the Methods section. In total, 18 independent samples with a random occupation of the three chemical species over the sites of an *fcc* structure have been used as initial conditions for the MC simulations. A large number of independent samples was used to improve sampling accuracy, and to provide a sufficient number of independent stacking fault planes to achieve converged average values of the SFE, as described below. For each sample, MC simulations were run for a total of 3,600 steps, representing 20 swap trials per atom.

Based on these calculations, the most significant trends in the potential energy change during the simulation are shown in Fig. 2A, using four representative samples. Although the number of swap trials per atom is small compared to classical MC simulations, the simulations produce SRO consistent with experiment and previous simulations, as described below. Importantly for the current study, these simulations provide configurations with varying states of chemical SRO, to enable a systematic investigation of the effect of local chemical ordering on the SFE and on the structural energy differences between the *fcc* and *hcp* phases. To quantitatively monitor the trend of local chemical ordering as well as its correlation with properties, four groups of configurations, *CH_0*, *CH_1*, *CH_2* and *CH_F*, were selected for detailed investigation according to the specific energy states (as denoted in Fig. 2A). *CH_0* and *CH_F* are the initial and final configurations from the simulation, while *CH_1* and *CH_2* are two intermediate states. Thus, each group contains 18 independent configurations, which are presented with features of local structural ordering and used to obtain averaged values in the following analysis

To describe the trends in local chemical ordering obtained in the MC simulations, we employ a parameter called the nonproportional number of local atomic pairs, $\Delta\delta_{ij}^{k}$, where *i* and *j* are the type of atom species, *k* denotes the $k^{\text{th}}$ nearest-neighbor shell (see Methods). This parameter, which is related to the conventional Warren-Cowley parameters used in alloy theory (24), describes the preferred/unfavored pairing around an atomic species within the first, second and third nearest-neighbor shells (see the illustration of atomic configurations in Fig. 2B), whose coordination numbers are 12, 6 and 24, respectively. Positive (negative) values of $\Delta\delta_{ij}^k$ correspond to a tendency of unflavored (favored) $i$-$j$ pairs. In the case of a random solution, $\Delta\delta_{ij}^k = 0$.

Figure 2C plots the development of $\Delta\delta_{Cr-Cr}$ in CrCoNi alloys for the first three nearest-neighbor shells during the MC simulation (the number of Cr-Cr pairs are found to deviate most significantly from a random solid solution, as compared to the other five pair types). Following the trend in the energy shown in Fig. 2A, the evolution in the local chemical order is initially rapid, but then becomes more gradual at the later stages of the simulation. As is commonly observed for chemical short-range ordering in alloys, contrasting behavior is obtained for nearest-neighbor and further-neighbor shells; as shown in Fig. 2C, the number of Cr-Cr pairs is reduced in the first nearest-neighbor shell of Cr atoms, while the second and third nearest-neighbor shells show an enhancement.

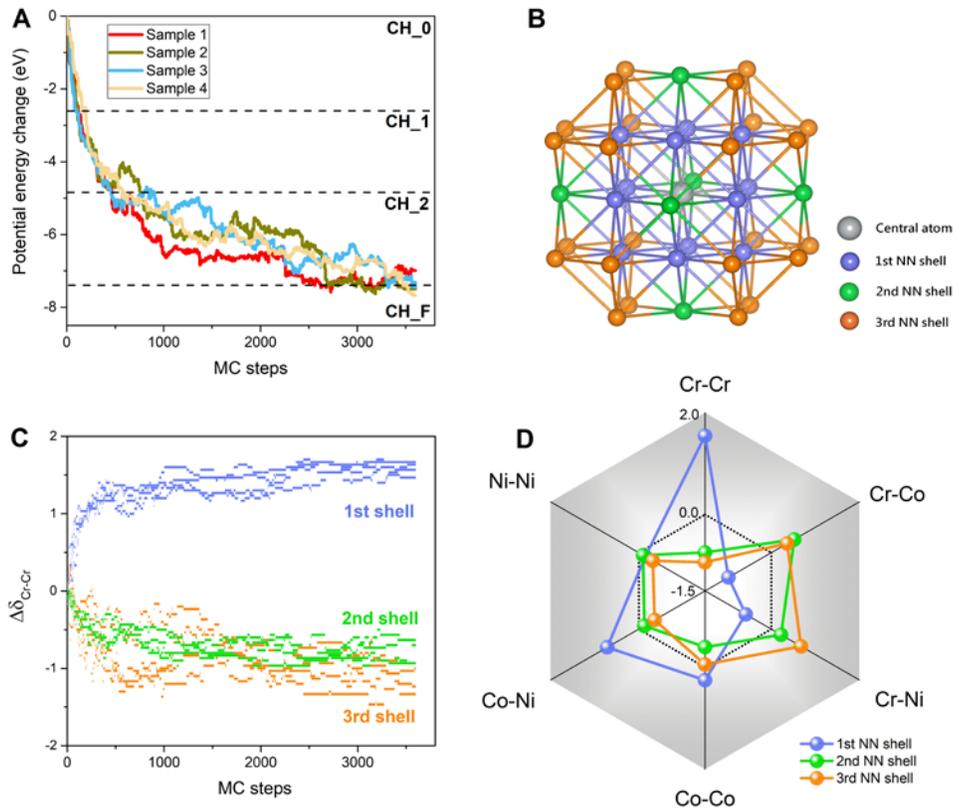

**Figure 2. Evolvement of energy states and local chemical ordering in CrCoNi alloys.** The CrCoNi ternary alloys (a total of 18 independent samples) were studied by DFT-based MC simulations (see Methods). (A) Potential energy changes with the steps of the MC simulation. Only data for four

representative samples are plotted in this figure. Configurations at the specific energy states are denoted into four groups as: *CH_0*, *CH_1*, *CH_2* and *CH_F*. (B) Schematic illustration of the first, second and third nearest-neighbor (NN) shells in a *fcc* crystal; (C) the nonproportional number of local Cr-Cr pairs, $\Delta\delta_{Cr-Cr}$, in the first three NN shells are involved with the steps of MC simulation; (D) Averaged over 18 independent CrCoNi alloys at final state (*CH_F*), the nonproportional number of local atomic pairs, $\Delta\delta_{ij}^k$, for Cr-Cr, Cr-Co, Cr-Ni, Co-Co, Co-Ni and Ni-Ni are plotted at the first, second and third nearest-neighbor shells. The dashed lines represent the case of a random solid solution.

For the final state of the MC simulation (*CH_F*), the local chemical order, averaged over the 18 independent samples, is plotted in Fig. 2D, in terms of the values of $\Delta\delta_{ij}^k$ for *ij*=Cr-Cr, Cr-Co, Cr-Ni, Co-Co, Co-Ni and Ni-Ni pairs at *k*=first, second and third nearest-neighbor shells. The dashed lines represent the case of a random solid solution. Deviations from random behavior are clearly evident. As already indicated in Fig. 2C, the most pronounced local chemical ordering is around Cr atoms. Specifically, at the short-range scale, $\Delta\delta_{Cr-Cr}^1$ is about 1.58, while $\Delta\delta_{Cr-Co}^1 \approx -0.98$ and $\Delta\delta_{Cr-Ni}^1 \approx -0.61$, as seen in Fig. 2D. The general trend related to chemical order for Cr-Cr pairs is consistent with previous DFT-MC simulations (20), and also with EXAFS analysis (21), as described above. The chemical order around Cr atoms is still significant for the second and third nearest-neighbor shells, but with the opposite trend to that for nearest-neighbors (*i.e.*, an enhancement in Cr atoms and depletion in Ni and Co atoms around a central Cr, as shown in Fig. 2D). Similar trends are also observed for the chemical ordering around Co and Ni atoms

**Stacking fault energy of random solid-solution CrCoNi alloys**

In the DFT calculations of the SFE for both intrinsic stacking faults (*ISF*) and extrinsic stacking faults (*ESF*) as illustrated in Fig. 3A-D, we consider slabs containing 360 atoms created by sequentially stacking 12 close-packed (111) atomic planes in their regular *ABCABCABCABC* configuration. An *ISF* was created by shifting the top six layers along the $\langle 112 \rangle$ direction by the Burgers vector of the Shockley partial $b_s = \frac{a}{6}\langle 11\bar{2}\rangle$; this resulted in a stacking sequence: *ABCABC|BCABCA*. Further shifting of the fault leads to the *ESF*, as shown in Fig. 3C, resulting

in a stacking sequence of *ABCABC/B/ABCAB*. The value of $\gamma_{isf}$ and $\gamma_{esf}$ was obtained from the energy difference between the original and stacking-faulted structures, normalized by the corresponding area. Further details are given in the Methods section.

For the CrCoNi alloys with random compositional disorder, 108 independent as-assigned supercells were considered in total; the resulting calculated values of $\gamma_{isf}$ are shown in the inset to Fig. 3E. The calculated values of $\gamma_{isf}$ display a large range, from -140 to 65 mJ.m$^{-2}$, with an average value of -42.9 mJ.m$^{-2}$, as shown in Fig. 3E. In addition, the average value of $\gamma_{esf}$ is about -27.8 mJ.m$^{-2}$, which is also negative, but larger than $\bar{\gamma}_{isf}$. By comparison, additional values of $\gamma_{isf}$ were computed using replicated configurations from the smaller random solid-solution models described in the previous section on MC simulations (see Methods for details of replication). For each replicated sample, all of the six close-packed atomic planes were considered, and again 18 independent samples have in total 108 *ISF*, with the corresponding $\bar{\gamma}_{isf}$ values also plotted in Fig. 3E. These replicated samples lead to values close to those for the as-assigned supercells. This result provides validation for our calculation of SFE using the replicated configurations, which forms the basis for the studies, described in the next section below, of the effects of chemical SRO on SFE.

The magnitude of $\bar{\gamma}_{isf}$ calculated here, -42.9 mJ.m$^{-2}$, differs from previously published values of -24 mJ.m$^{-2}$ (11), -62 mJ.m$^{-2}$ (17), -40 mJ.m$^{-2}$ (18) and -48 mJ.m$^{-2}$ (25) that were also calculated using supercell methods. We view these differences with our present value of -42.9 mJ.m$^{-2}$ as reflecting the need, as illustrated in Fig. 3E, for a large number of samples to achieve well-converged averaged results (only limited configurations were considered in refs. [11,17,25]). Other DFT calculations, using the coherent potential approximation (CPA) or axial interaction model, yielded SFE values of -26 mJ.m$^{-2}$ (26) and -25 mJ.m$^{-2}$ (17), respectively. Despite these quantitative differences, the finding of *negative* $\bar{\gamma}_{isf}$ values for CrCoNi alloys with *random* configurational disorder is robust across all previous DFT calculations. As described above, this finding is at odds with experimental measurements, clearly warranting further examination. In the next section, the effect of local chemical ordering is considered, with the main finding that such chemical SRO is capable of leading to a change in sign of the SFE.

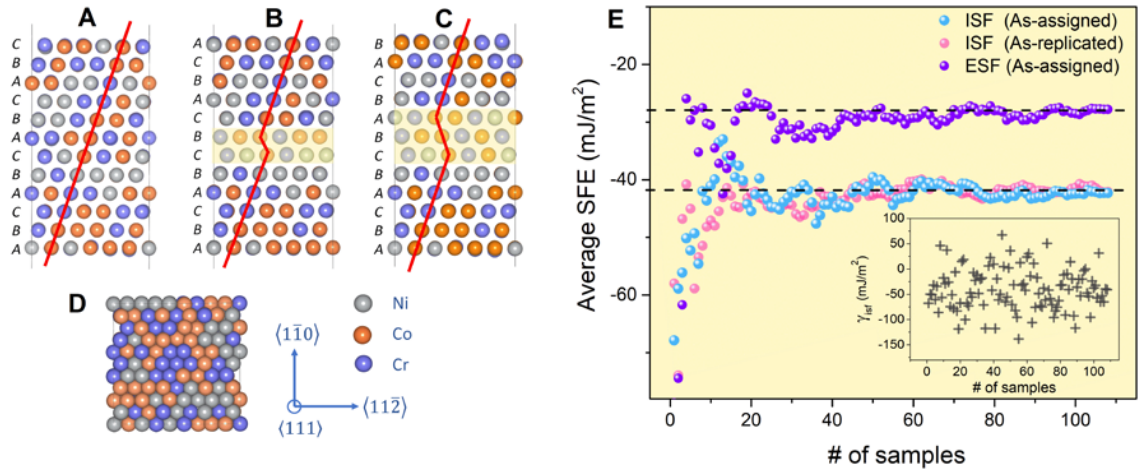

**Figure 3. Calculated stacking fault energy of the random solid-solution CrCoNi alloys.** (A-C) show the side view images of atomic configurations in the original *fcc* structure, with *ISF* and *ESF*, respectively; the yellow shade indicates the stacking fault; the plane labels of *ABC*... represent the sequence of close-packed (111) plane; (D) Top view image showing the close-packed (111) plane. (E) The average SFE with the number of random solid-solution CrCoNi samples (containing 360 atoms each) for both as-assigned and replicated supercells (see details in Methods). Inset shows the $\gamma_{isf}$ of as-assigned supercells of CrCoNi alloys.

**Stacking fault energy strongly correlates with local chemical order**

The inset in Fig. 3 shows a wide range of calculated $\gamma_{isf}$ values for random CrCoNi solid solutions, which is in contrast to a single value for a pure or perfectly-ordered crystalline alloy. A similar broad distribution can also be observed for CrCoNi alloys with chemical SRO (Fig. 4A). As plotted in Fig. 4A, the statistical variations of the calculated $\gamma_{isf}$ for four groups of configurations, *i.e., CH_0*, CH_*1*, CH_*2* and *CH_F*, all follow Gaussian type distributions. The mean value for CrCoNi alloys in *CH_0*, CH_*1*, CH_*2* and *CH_F* have been computed to be -42.9 mJ.m$^{-2}$, -14.9 mJ.m$^{-2}$, 13.3 mJ.m$^{-2}$ and 30.0 mJ.m$^{-2}$, respectively. The results establish a trend of increasing values of the SFE with increasing degree of chemical SRO. In contrast, the widths of the distributions (with the standard deviation of ~31 mJ.m$^{-2}$) are almost unchanged for varying states of chemical SRO. Although the exact width of the distributions depends on the size of the supercell used in the calculations, the results clearly illustrate how variations in the local composition and order affect planar fault energies. We note that a heterogeneous distribution of "local" SFEs may affect the plastic deformation mechanisms, considering the existence of both "weak" and "strong" close-packed atomic planes that is in contrast with conventional crystalline alloys. A similar example can be found in studying the dislocation cross-slip affected by the fluctuations in the spatial solute distribution in random solid solutions (27).

The average SFE, $\bar{\gamma}_{isf}$ and $\bar{\gamma}_{esf}$ in CrCoNi alloys can be quantitatively correlated with the degree of chemical SRO, reflected by the total nonproportional number of local atomic pairs, $\Delta\delta_{sum}$, as shown respectively in Fig. 4B-C. The $\Delta\delta_{sum}$ values, reflecting chemical order summed over all pairs (see Methods), are found to exhibit a monotonically linear relationship with the value of $\bar{\gamma}_{isf}$ and $\bar{\gamma}_{esf}$ (see Fig. 4B-C), for the first, second and third nearest-neighbor shells. By increasing the degree of local chemical order, the value of $\bar{\gamma}_{isf}$ can be tuned remarkably from an average value of -42.9 mJ.m$^{-2}$ to 30 mJ.m$^{-2}$, while the value of $\bar{\gamma}_{esf}$ can be tuned from -27.8 mJ.m$^{-2}$ to 66 mJ.m$^{-2}$. Such variations, in *$\bar{\gamma}_{isf}$ by more than 70 mJ.m$^{-2}$, and $\bar{\gamma}_{esf}$ by more than 90 mJ.m$^{-2}$*, are substantial for a crystalline metal or alloy and would be expected to significantly impact mechanical deformation mechanisms, as described in the Discussion section below.

It is also worth noting that the experimental measured $\gamma_{isf}$ is 22±4 mJ.m$^{-2}$ at 300 K, which already falls within the range explored by our DFT calculation at 0 K in Fig. 4B. In comparing our zero-temperature calculated values with experiment measurements at room temperature, it is important to note that the vibrational contributions to the stacking-fault free energies need to be considered at finite temperature (18,28). In recent work by Niu *et al*. (18), such vibrational free energies were incorporated in the calculation of *hcp-fcc* free energy differences for random CoCrNi alloys, with the finding that the free energies changed by less than 15% in the temperature range between 0 K and 300 K. We thus expect that the vibrational contributions to the SFE values will be similarly relatively smaller over this temperature range in such alloys, and the measured values for SFEs in experimental CrCoNi samples extrapolated to 0 K would be slightly smaller than the reported room temperature value of 22±4 mJ.m$^{-2}$. *This finding appears to imply that the CrCoNi alloys that have been investigated experimentally to date likely possess a significant degree of local chemical SRO, which falls within the range investigated by our DFT calculation, rather than being a random solid solution.*

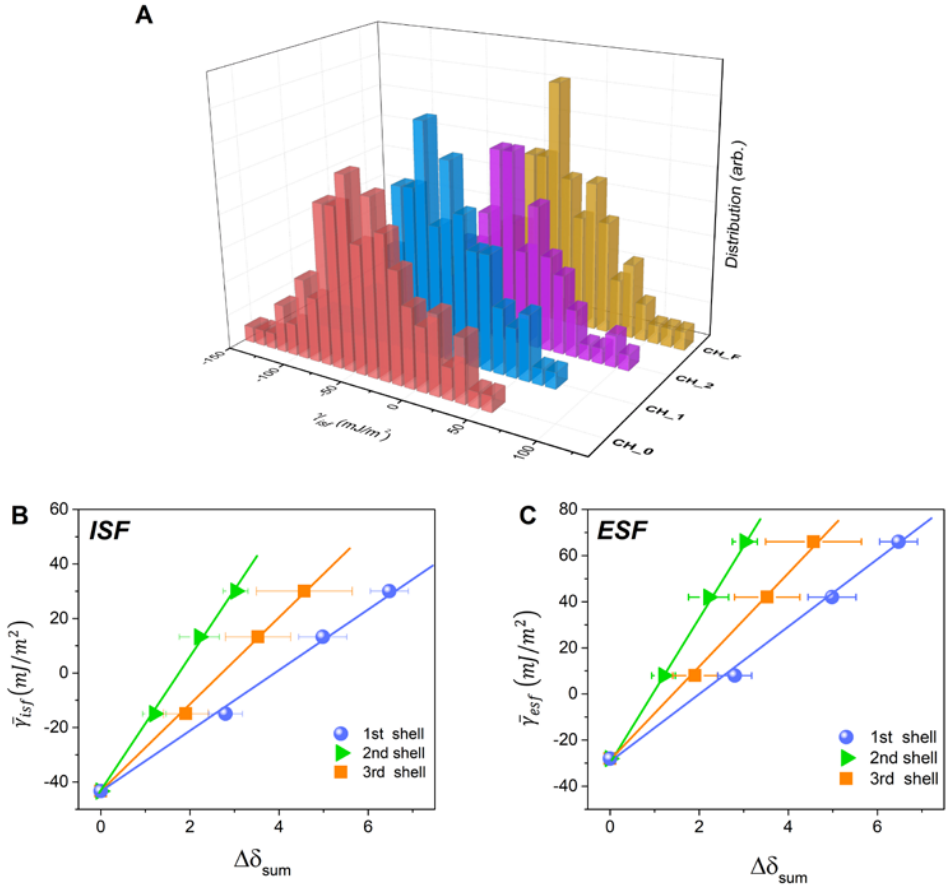

**Figure 4. Stacking fault energy correlates strongly with local chemical ordering.** (A) Distribution of intrinsic stacking fault energy, $\gamma_{isf}$, for the CrCoNi alloys in four specific states, *i.e.*, *CH_0*, *CH_1*, *CH_2* and *CH_F*, which span from random solid solution to the highest degree of chemical ordering. Totally 108 stacking faults were considered for analysis in each group. The average energy, $\bar{\gamma}_{isf}$ and $\bar{\gamma}_{esf}$ in (B) and (C), respectively, among those four groups were correlated with the total nonproportional number of local atomic pairs, $\Delta\delta_{sum}$ for the first, second and third nearest-neighbor shells.

## Occurrence of *fcc* versus *hcp* phase in CrCoNi solid-solution alloys

Understanding the competition of *fcc* and *hcp* phases is critical for CrCoNi solid-solution alloys. Firstly, it is expected that the SFE in *fcc* crystal correlates with the value of $\Delta E^{fcc \to hcp}$, as the intrinsic stacking fault can be considered as a two-layer embryo of the *hcp* structure embedded in the *fcc* matrix (see ref. 29 by Olson and Cohen). Secondly, the as-prepared CrCoNi medium-entropy alloys crystallize as *fcc* solid solutions in experiment, while this is seemingly in conflict with previous computational results showing that the *hcp* phase is energetically favored relative to the *fcc* phase (11,13,18). Moreover, at large strain levels of deformation in CrCoNi alloys, a new phase with *hcp* lamellae has been reported to appear within the matrix of the *fcc* phase (13). This could be suggestive of a transformation-induced plasticity effect that would be expected to occur if the *hcp* structure has a lower energy than that of the *fcc* phase, which then would be considered to be metastable. More recently, Niu *et al.* revealed the magnetic contribution to the relative phase stability and unique dislocation mechanisms for *fcc* to *hcp* phase transformation, by using DFT calculations (18). In light of these considerations, it is very important to investigate the effects of local chemical order on the *hcp-fcc* energy difference ($\Delta E^{fcc \to hcp} = E_{hcp} - E_{fcc}$).

Figures 5A-B show the side and top view images, respectively, of atomic configurations of CrCoNi alloys in *fcc* and *hcp* structures. The sequence of close-packed (111) planes changes from *ABCABC* for *fcc* phase to *ABABAB* for *hcp* phase. The relative energies for those two phases in CrCoNi alloys are plotted in Fig. 5C, for four specific states of chemical order, *i.e.*, *CH_0*, *CH_1*, *CH_2* and *CH_F*. The green shades indicate the corresponding values of $\Delta E^{fcc \to hcp}$. For the random solid solution (*CH_0*), the *hcp* phase is energetically favored. On increasing the degree of local chemical order, the energies for both phases decrease, but the energy of the *fcc* phase drops faster, as seen in Fig. 5C. At the final state (*CH_F*), the value of $\Delta E^{fcc \to hcp}$ reverses, favoring the *fcc* phase. The behavior of $\Delta E^{fcc \to hcp}$ can be quantitatively correlated with $\Delta \delta^1_{Cr-Cr}$ so that it reflects the local chemical order, as shown by the inset in Fig. 5C.

Figure 5D plots $\bar{\gamma}_{isf}$ versus $\Delta E^{fcc \to hcp}$ for CrCoNi alloys with varying chemical SRO and a nearly linear relation is obtained, as theoretically predicted. For the random solid solution, the DFT-calculated negative value of $\Delta E^{fcc \to hcp}$ correlates with the negative value of $\bar{\gamma}_{SFE}$. In comparison, for the highest degree of chemical SRO, the *fcc* CrCoNi alloys becomes more stable than the *hcp* phase, which corresponds to a $\bar{\gamma}_{SFE}$ value as high as ~30 mJ.m$^{-2}$. Dashed lines in Fig. 5D indicate where $\bar{\gamma}_{isf} = 0$ or $\Delta E^{fcc \to hcp} = 0$. Of note is where the *hcp* phase is energetically equivalent to the *fcc* phase, the corresponding $\bar{\gamma}_{SFE}$ is predicted to be approximately 18 mJ.m$^{-2}$ (two times of the interfacial energy between the *fcc* and *hcp* phases (29)). In particular, there are states of chemical order that give rise to positive stacking-fault energies (consistent with experimental measurements of the SFEs in ref. 9), but with very low or even negative values of $\Delta E^{fcc \to hcp}$. For such alloys, deformation may enable the *in situ* transformation of *fcc* to *hcp* structures, which can contribute to the steady high strain-hardening rates (13), which in turn are likely to result in both high tensile strength and ductility in such alloys, as discussed in the Discussion section below.

Therefore, two critical results have been collected for CrCoNi solid-solution alloys: i) Tuning the local chemical order can significantly change the preference between *fcc* and *hcp* phases, from negative $\Delta E^{fcc \to hcp}$ to the positive value; ii) The quantitative relationship between the $\Delta E^{fcc \to hcp}$ and SFE can be established for this alloy at various states of local chemical order. Those findings can shed new insights to understand the phase stability as well as deformation-induced phase transformation in CrCoNi solid-solution alloys.

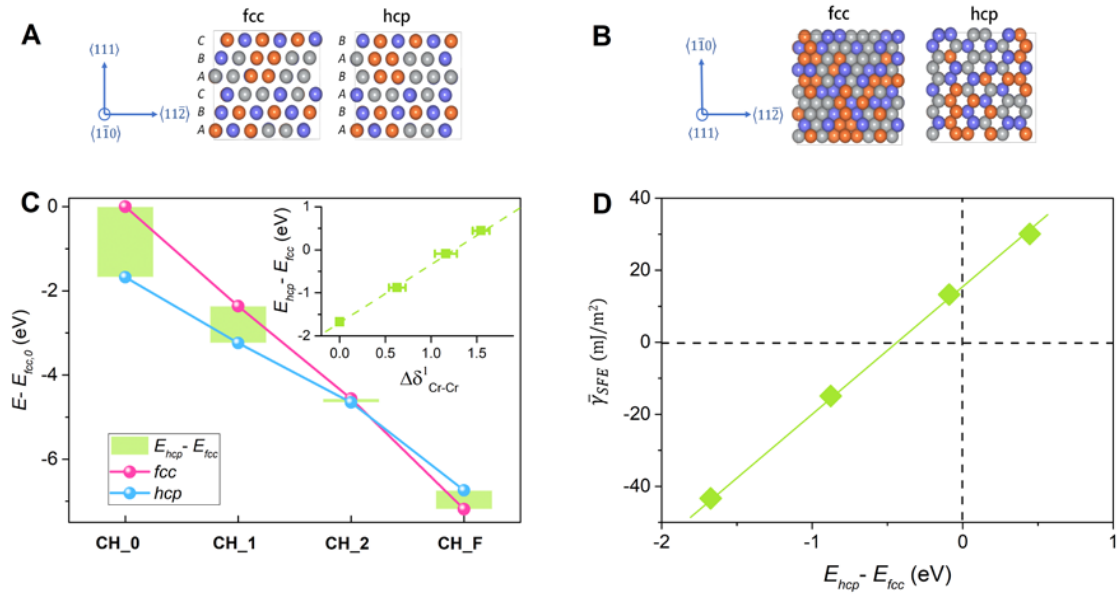

**Figure 5. The *fcc* versus *hcp* phase in CrCoNi solid-solution alloys.** (A) The side and (B) top view images of atomic configurations for the CrCoNi alloys in both *fcc* and *hcp* phases. (C) The relative energy difference ($E$-$E_{fcc,0}$) for both *fcc* and *hcp* phase at various degrees of local chemical ordering, where $E_{fcc,0}$ is the average potential energy of the *fcc* CrCoNi alloys with random solid solution; the inset shows the correlation between $\Delta \delta^1_{Cr-Cr}$ and energy difference between the *hcp* and *fcc* phases, $\Delta E^{fcc \to hcp}$; (D) The correlation between the average stacking fault energy, $\bar{\gamma}_{isf}$, and $\Delta E^{fcc \to hcp}$. Dashed lines indicate where $\bar{\gamma}_{isf} = 0$ or $\Delta E^{fcc \to hcp} = 0$.

**Discussion and Conclusions**

Stacking fault energies and structural energy differences in CrCoNi solid solutions, as represented by medium-entropy alloys and as a basis for the Cantor CrMnFeCoNi high-entropy alloy, have been studied in this work using DFT calculations, with a specific focus on the effect of local chemical ordering. From MC simulations, we find strong tendencies for the formation of chemical SRO, particularly around Cr atoms, which favor Ni and Co atoms as nearest neighbors, and an enhancement of Cr neighbors in the second and third shells. It is important to emphasize that the state of local chemical order in CrCoNi solid solutions, as well as other HEAs, need not be that corresponding to an equilibrium thermodynamic state, as it is inherited through cooling from high temperatures. This is very significant as it means that not only the microstructure, but also the local chemical order can be controlled, *e.g.*, by adjusting the heat treatment at elevated temperatures, to enhance or disfavor the degree of ordering. However, this has yet to be proven experimentally, in part because of the difficulty in assessing regions of chemical order. *However, the strong effects of local chemical order on the underlying planar-fault and structural energy differences found in this theoretical study emphasize that experimental studies of structure-properties relationships in these multiple-element materials should now involve additional detailed measurements of chemical order.* Indeed, this is becoming increasingly feasible through the use of atom probe tomography and advanced transmission electron microscopy techniques (30,31).

The present calculations, summarized in Fig. 4, demonstrate that the average SFE of CrCoNi solid-solution alloys can vary remarkably as the degree of chemical SRO increases. Specifically, as the degree of chemical SRO is varied from the random configuration up to the highest degree of order realized in the MC simulations, the value of $\bar{\gamma}_{isf}$ increases from -42.9 mJ.m$^{-2}$ to 30 mJ.m$^{-2}$, while the value of $\bar{\gamma}_{esf}$ ranges between -27.8 mJ.m$^{-2}$ to 66 mJ.m$^{-2}$. In comparison, the experimentally measured value of $\bar{\gamma}_{isf}$, 22±4 mJ.m$^{-2}$, falls within the range explored by our DFT calculation, implying that CrCoNi alloys may contain a degree of local chemical order, although not necessarily the value corresponding to thermodynamic equilibrium at room temperature (since it is inherited from cooling from high temperature). This notion offers a profound resolution to the discrepancy between experimental SFE measurements and previous DFT calculations that have found negative values for the SFE of this alloy. Specifically, in all the DFT calculations of the SFE of CrCoNi alloys prior to this work, chemical ordering has not been explicitly accounted for.

The present computational study has revealed four characteristic features of the stacking fault energies in medium-entropy CrCoNi solid-solutions: i) The average SFE is highly tunable with a substantial variation of more than 70 mJ.m$^{-2}$ for $\bar{\gamma}_{isf}$, and more than 90 mJ.m$^{-2}$ for $\bar{\gamma}_{esf}$ by tailoring local chemical ordering; ii) The low SFE of CrCoNi solid-solution alloys serves to promote the formation of deformation nanotwins, which has been associated with the development of high strength, ductility and toughness in crystalline alloys; iii) This tunable SFE, which correlates with the negative or positive value of $\Delta E^{fcc \rightarrow hcp}$, can also affect deformation-induced phase transformation from *fcc* to *hcp* phase; iv) The predicted SFEs are found to exhibit a large distribution in values that would be realized in an alloy solid solution, where the state of local chemical order varies heterogeneously throughout the sample. These unique aspects of stacking faults in CrCoNi solid solutions, resulting from the existence and variation of local chemical ordering, would be expected to influence the mechanisms of plastic deformation, and hence the mechanical properties. Further details of SFE effect in conventional crystalline alloys are discussed in the Supplementary Information. Compared to conventional alloys, however, the consequences of the unique aspects of planar fault energies for macroscopic deformation behavior remains incompletely understood for multicomponent HEAs. Thus, we believe that it is essential that further systematic studies be conducted on local chemical order as well as its role in influencing the mechanical deformation of HEAs in the future.

The findings from the present work highlight the general need to discern how critical the role of local chemical ordering is on the properties of other HEA systems (2,32-35). It is important to note here that chemical order may not only affect the energies of stacking faults, but also other critical defects, such as vacancies, interstitials, and twin boundaries, all of which could be similarly affected by tailoring the local chemical order to influence macroscopic mechanical properties. Strong variations in these defect energies will likely affect not only strength, ductility and toughness, but also resistance to high-temperature creep and even irradiation damage (36-38). Future research on structure-property relationships in these alloys should thus include consideration of the effects of local chemical order, to understand the degree to which chemical SRO can be used as an independent structural variable to guide alloy design and optimization. Our results thus highlight the possibility of "tuning order in disorder" to ultimately achieve the science-based design and optimization of new HEA systems with specifically desired combinations of macroscale mechanical properties.

**Methods**

**DFT-based Monte Carlo simulations:** The systems studied were equimolar CrCoNi ternary alloys in the *fcc* phase with lattice constant of 3.526 Å. For the Monte Carlo simulation, 18 independent supercells, containing 180 atoms each, were generated as special quasi-random structure (SQS) for initial starting points; SQS structures represent the best representative model of a random alloy in small systems with periodic boundary conditions (39). The temperature employed in the MC simulations was 500 K. Energy calculations were performed with the Vienna *ab initio* simulation package (VASP) (40-42), using spin-polarized density-functional theory, with a plane wave cut-off energy of 420 eV. Brillouin zone integrations were performed using Monkhorst–Pack meshes with a 2×2×2 grid (43). Projector-augmented-wave (PAW) potentials (42,44) were employed with the Perdew-Burke-Ernzerhof (PBE) generalized-gradient approximation (GGA) for the exchange-correlation functional (45). In the DFT simulations the magnetic moments were initialized in a ferromagnetic state, although the final spin configurations were disordered and varied with the state of chemical order. Similar to the methods utilized in ref. 20, lattice MC simulations included swaps of atom types with the acceptance probability based on the Metropolis–Hastings algorithm (46). Finally, a conjugate-gradient algorithm was employed for structural relaxations at the end of the MC runs using a denser *k*-point mesh of 3×3×3.

**Local chemical order parameter:** Modified from the Warren-Cowley parameter (24), we defined the nonproportional number of local atomic pairs, $\Delta\delta_{ij}^{k}$, to quantify the chemical ordering around an atomic species for the first, second and third nearest-neighbor shells, for which the corresponding coordination numbers are 12, 6 and 24 respectively. $\Delta\delta_{ij}^{k}$ was calculated as:

$$\Delta\delta_{ij}^{k} = N_{0,ij}^{k} - N_{ij}^{k} \quad , \tag{1}$$

where *k* denotes *k*th nearest-neighbor shells, $N_{ij}^{k}$ is the actual (average) number of pairs between atoms of type *j* and type *i* in the $k^{\text{th}}$ shell and $N_{0,ij}^{k}$ is the number of pairs that are proportional to the corresponding concentrations. In the case of a random solution $\Delta\delta_{ij}^k = 0$. The total nonproportional number of all atomic pairs in the $k$th shell, $\Delta\delta_{sum}^k$, was defined as:

$$\Delta\delta_{sum}^k = \sum_i \sum_j \left|\Delta\delta_{ij}^k\right| \quad . \tag{2}$$

**Stacking fault energy calculations:** The average SFE, $\bar{\gamma}_{isf}$ and $\bar{\gamma}_{esf}$ of CrCoNi alloys, has been calculated using supercell models, employing the same DFT methods described above. All the SFE values were calculated using 360-atom supercells, as illustrated in Fig. 3A-C, which were set up by sequentially stacking 12 close-packed (111) atomic planes. Such large supercells were used to reduce the size effect on the calculated values of SFE. Two cases of sample initialization were implemented. The first case was constructed by replicating the 180-atom configurations in $z$ direction normal to the stacking-fault plane, including those at four different states, *CH_0*, *CH_1*, *CH_2*, and *CH_F*; for each state, there were 108 stacking faults under study, extracted from 18 independent samples, whose six planes were all considered for each configuration by shifting the stacking fault to the middle of the constructed supercells. The second case was for the directly as-assigned 360-atom configuration with random solid solution, representing 108 independent configurations in total. The atomic configurations for each sample were structurally relaxed in the three directions. Next, an intrinsic stacking fault was created by shifting the top six layers along the $\langle 112 \rangle$ direction by the Burgers vector ($b_s$) of the Shockley partial $b_s = \frac{a}{6}\langle 11\bar{2}\rangle$, where $a$ is the lattice parameter; this resulted in a stacking sequence: *ABCABC/BCABCA*, as shown in Fig. 3B. Further shifting of the fault leads to the *ESF*, as shown in Fig. 3C, resulting in a stacking sequence of *ABCABC/B/ABCAB*. The SFE was obtained from the energy difference of the two optimized structures normalized by the area of the stacking fault.

**Code availability:** The computer code used in this study is available from the corresponding author (M. Asta) upon request.

**Acknowledgements**

This work was supported by the Mechanical Behavior of Materials Program (KC13) at the Lawrence Berkeley National Laboratory, funded by the U.S. Department of Energy, Office of Science, Office of Basic Energy Sciences, Materials Sciences and Engineering Division, under Contract No. DE-AC02-05CH11231. The study made use of resources of the National Energy Research Scientific Computing Center, which is also supported by the Office of Basic Energy Sciences of the U.S. Department of Energy under Contract No. DE-AC02-05CH11231.

**Author contributions**

J.D., M.A. and R.O.R designed the research; J.D. performed the simulations; J.D., M.A. and R.O.R. analyzed the data; and J.D., Q.Y., M.A. and R.O.R wrote the paper.

**Data availability**

Data can be made available on request to the corresponding authors, specifically Mark Asta at mdasta@lbl.gov.

**Competing interests**

The authors declare no competing interests, financial or otherwise.

Supporting Information for

# Tunable stacking fault energies by tailoring local chemical order in CrCoNi medium-entropy alloys

**Jun Ding**[1]**, Qin Yu**[1]**, Mark Asta**[1,2*]**, Robert O. Ritchie**[1,2†]

[1] Materials Sciences Division, Lawrence Berkeley National Laboratory, Berkeley, CA94720, USA
[2] Department of Materials Science & Engineering, University of California, Berkeley, CA 94720, USA
Corresponding author: email: * mdasta@berkeley.edu (M.A.); † roritchie@lbl.gov (R.O.R)

**SI Text**

**Effects of stacking fault energy on plastic deformation mechanisms in conventional crystalline alloys**

We have found in this study that a substantial variation in intrinsic and extrinsic stacking fault energy (SFE), in excess of 70 mJ.m$^{-2}$ and 90 mJ.m$^{-2}$, respectively, can be achieved by tailoring the local chemical SRO in medium-entropy CrCoNi solid-solution alloys. Similarly, large variations in SFE can also be achieved by controlling solution concentration (or alloy composition) in conventional *fcc* alloys, such as Cu-Al [47], Ni–Cu [48], Ni–Fe [48], Ni-Co [49], Co-Ni-Cr-Mo [49], and Fe-Mn based (TWIP/TRIP steel) alloys [15,50,51], although their SFEs are invariably always positive. Theoretically, lowering the SFE in *fcc* alloys makes it easier for perfect dislocations to be dissociated into two Shockley partials with a wide stacking fault ribbon [52]. This process reduces the ability of the screw components of dislocation to cross-slip onto other slip planes, thus restricting dislocation slip to a planer mode [52]. Cell-like dislocation configurations are formed as a consequence of wavy slip in high SFE metals, whereas planar arrays (or coplanar) dislocation groups and wide stacking faults are often observed in low SFE metals [47].

A low intrinsic and extrinsic SFE in an alloy promotes the formation of twins, which is consistent with observations of extensive deformation nanotwin networks in CrCoNi solid-solution alloys [11]; as the formation of such twins is a locally stress-controlled process, deformation twinning is often favored at cryogenic temperatures. The preference for twinning over dislocation slip in low SFE metals is rooted in the physical process of twinning. The nucleation and growth of twins are associated with lattice shearing through successive glide of Shockley partial dislocations on consecutive close-packed atomic planes in *fcc* metals [53,54]. A low SFE sets a lower energy barrier for glide of twinning partials, thus leading to a lower twinning stress. Indeed, the enhanced formation of deformation nanotwinning in the equiatomic CrCoNi alloy has been associated with its exceptional damage-tolerant behavior [11]. Twin boundaries impose effective barriers to dislocation motion, leading to a "dynamic Hall-Petch" effect [15,51]. It is also important to note that strengthening by twin boundaries does not cause the loss of material ductility, while coherent twin boundaries have a unique character that allows for the glide of impinging dislocation along their interfaces [11,50,55,56]. Therefore, besides the shearing induced by twinning itself, the incorporation of incident dislocations on twin boundaries provides a mechanism for additional plasticity.

Lastly, the low SFE, which correlates with a low value of $\Delta E^{fcc \to hcp}$, can also favor the deformation-induced phase transformation from *fcc* to *hcp* phase. For example, the activation of such phase transformations has been found in conventional crystalline alloys with low SFEs, such as below 12~20 mJ.m$^{-2}$ for Fe-Mn based alloys [50], and below ~10-15 mJ.m$^{-2}$ for Co-Ni-Cr-Mo alloys [49]. Theoretically, this relates to the fact that the *fcc* to *hcp* phase transformation is related to dissociation of perfect dislocation into Shockley partials [15,51] as well as passage of one Shockley partial on every other close-packed atomic plane. In analogy to the twinning-induced strengthening effect, the deformation-induced *hcp* lamellae, as well as nanotwin-*hcp* lamellae, have been proposed to contribute to the higher strain hardening and tensile strength in the CrCoNi alloy when compared to the CrMnFeCoNi Cantor alloy [11]. The volume fraction of *hcp* structure increased with increasing plastic deformation; notably, the increase in the volume fraction of *hcp* structure at cryogenic temperature was much faster than at room temperature.